\documentstyle[emulateapj-djb,epsf,astroref-djb]{article}   

\tighten  \singlespace 

\def\aco{${\rm ^{12}CO}$}  \def\bco{${\rm ^{13}CO}$}
\def\c18o{${\rm C^{18}O}$} \def\h2o{${\rm H_2O}$}
\def\nh3{${\rm NH_3}$}
\def\sgra{Sgr A$^*$}

  \def\mic{\,\mu\rm m} 
\def\lesssim{\mathrel{\hbox{\rlap{\hbox{\lower4pt\hbox{$\sim$}}}\hbox{$<$}}}}
\def\moresim{\mathrel{\hbox{\rlap{\hbox{\lower4pt\hbox{$\sim$}}}\hbox{$>$}}}}
\def\lapprox{\;\raise.4ex\hbox{$<$}\kern-0.8em \lower .6ex\hbox{$\sim$}\;}
\def\gapprox{\;\raise.4ex\hbox{$>$}\kern-0.8em \lower .6ex\hbox{$\sim$}\;}
         \def\kms{\,\rm km\,s^{-1}} 
\def\cc{\,{\rm cm}^{-3}}      \def\cmi{\,\rm cm^{-2}}

\def\per{~}
\received{** 1999} \accepted{** 1999} \journalid{337}{15 January 1989}
\articleid{11}{14} \slugcomment{}

\begin{document}

\title{Cold H$_2$O and CO ice and gas toward the Galactic Center{\bf$^1$}}

\author{Andrea Moneti}
\affil{CSIC, IEM, Dpto. F\'{\i}sica Molecular, Serrano 121, E-28006
Madrid, Spain.\\
e-mail: amoneti@astro.iem.csic.es}
\author{Jos\'e Cernicharo}
\affil{CSIC, IEM, Dpto. F\'{\i}sica Molecular, Serrano 121, E-28006
Madrid, Spain.\\
Division of Physics, Mathematics and Astronomy, California Institute
of Technology,\\ MS 320-47, Pasadena, CA 91125, USA\\
e-mail: cerni@astro.iem.csic.es}
\author{Juan Ram\'on Pardo}
\affil{Division of Physics, Mathematics and Astronomy, California Institute
of Technology,\\ MS 320-47, Pasadena, CA 91125, USA\\
e-mail: pardo@submm.caltech.edu}

          \footnotetext[1]{Based on observations with ISO,
          an ESA project with instruments funded by ESA Member States
          (especially the PI countries: France Germany, the Netherlands
          and the United Kingdom) and with the participation of 
          ISAS and NASA.}

\begin{abstract}

We present observations of CO, $^{13}$CO and of H$_2$O in the middle and
far-infrared taken with the ISO-SWS and ISO-LWS spectrometers toward two
positions in the Galactic Center region (Sgr A$^*$ and GCS-3).  Both ice
and gas phase molecules are detected.  The ISO data have been complemented
with observations of the J=3-2 and J=7-6 lines of CO carried out at the
Caltech Submillimeter Observatory. The ISO and CSO data indicate that the
absorbing gas is extremely cold, T$_K \simeq 10$ K, suggesting that it is
located in the dark clouds of the different spiral arms that intersect the
line of sight towards the Galactic Center.  From the analysis of the CO
absorption we derive $^{13}$CO gas phase column densities of 1.1 and
$0.7\per 10^{17}\cmi$ towards \sgra\ and GCS-3 respectively.  The H$_2$O
gas column density in the direction of Sgr A$^*$ is $\simeq 2\per
10^{16}\cmi$. The derived CO/\h2o and gas/solid abundance ratios
corresponding to these cold clouds are remarkably similar along the two
lines of sight. We find that nearly all the CO is in the gas phase, while
the \h2o is almost entirely frozen onto the surfaces of cold dust grains.
Finally, the N$_{gas+ice}$(CO)/N$_{gas+ice}$(H$_2$O) abundance ratio is
$\simeq 5$ implying that H$_2$O formation processes are highly efficient.

\end{abstract}

\section{Introduction} 

The water molecule plays an important role in the chemistry of interstellar
and circumstellar clouds. However, due to its high abundance in the
terrestrial atmosphere, its rotational, vibrational, and electronic
transitions of the main isotopic species are mostly hidden from view from
ground-based telescopes (except for some weak rotational lines, see
Cernicharo et al. 1990, 1994).  It was only with the advent of the Infrared
Space Observatory (ISO, Kessler et al., 1999) that extensive observations
of the far-infrared thermal lines of water vapor could be used to obtain
reliable estimates of the water vapor abundance in molecular clouds.
Mapping of the Sgr B2 molecular cloud with ISO by Cernicharo et al. (1997a)
has definitely shown that H$_2$O is a ubiquitous molecule in the cores of
warm molecular clouds with an abundance of $\simeq 10^{-5}$.  In addition,
maps of the emission of several H$_2$O lines in Orion IRc2 have been
obtained by Cernicharo et al. (1997b, 1999) showing that the abundance of
H$_2$O is close to 10$^{-4}$ in the shocked gas.  H$_2$O observations of
the central position of Orion have been also obtained by van Dishoeck et
al.  (1998), Gonz\'alez-Alfonso et al. (1998) and Harwit et al. (1998).
All these new data refer to the warm regions of molecular clouds.  While
there is also strong evidence for a high water depletion onto the ice
mantles of dust grains in the coldest regions, no water vapor has been
detected in the direction of dark clouds, probably due to the low
temperature and density prevailing in these objects.

In order to study the coldest and least disturbed regions of molecular
clouds, we have examined the ISO archive spectra of two positions in the
Galactic Center, where absorption from the intervening medium is known to
occur (Geballe, Baas, and Wade 1989, Lutz et al.~1996, Okuda et al.~1990).
In the case of \sgra, this cold gas is also seen in the form of sharp and
deep absorption features at 0, $-30$, and $-50\kms$ in various millimeter
transitions of \nh3\ and CO against the broad background produced by the
warm clouds in Sgr A West (Serabyn and G\"usten 1986, Sutton et al.~1990).
These data, together with submillimeter observations of the J=3-2 and J=7-6
lines of CO, have been modeled to place constraints on the abundances of
these molecules in both the gas and the solid phase.

\section{Observations}

All data were obtained with the ISO Short and Long Wavelength Spectrometers
(SWS, de Graauw et al. 1999; LWS, Clegg et al., 1999) using the full
resolution grating mode of the SWS ($\lambda/\Delta\lambda \simeq
1500$--2000) and the grating and Fabry Perot (FP) modes of the LWS
($\lambda/\Delta\lambda \simeq300$ and 9000 respectively).  The effective
SWS aperture in the 4--$7\mic$ range is about $14''\times 20''$, while the
LWS aperture is circular and $\sim 70''$ in diameter.  In the case of
GCS-3, the aperture was centered on GCS-3 I, though the brightest source in
the aperture at $5\mic$ is GCS-3 II (Moneti et al.~1999).  The aperture
centered on \sgra\ also includes several point-like and diffuse sources.
OLP 7.1 archive data were further processed with the ISAP
package\footnote{ISAP is a joint development of the LWS and SWS instrument
teams and data cent-res.  Contributing institutes are CNRS, IAS, IPAC, MPE,
RAL, and SRON.}.  The results are shown in Figures 1, 2 and 3. In order to
resolve the velocity structure of the absorbing/emitting gas in the
direction of Sgr A$^*$ we have made observations of the J=3-2 and J=7-6
lines of CO in the direction of this source with the CSO telescope. These
observations were made in position switching, with the off position 4
degrees outside the galactic plane. A region of 10x6 arcminutes was mapped
in the J=3-2 line of CO and a few positions were observed in the J=7-6 line
of the same species.  The system temperatures at the frequencies of the
J=3-2 and J=7-6 lines of CO were 760 and 6000 K respectively and the
spectral resolution was 0.5 MHz. The results for the central position of
our maps, Sgr A$^*$, are shown in Figure 3.

\section{Results and Discussion}

In order to determine the physical conditions of the foreground cold gas
from the ISO H$_2$O and CO data, we have used an ETL model with line
parameters from the HITRAN molecular database (Rothman et al., 1993) to
find the best fit to the line absorption.  We note, however, that the
energies of all ortho levels of H$_2$O are referred in HITRAN to the ground
state of para-H$_2$O, i.e., the energy of the $1_{01}$ ground state of
ortho-H$_2$O is not set equal to zero.  For very low rotational
temperatures the derived physical parameters will have a large error if the
HITRAN values of the energies and line intensities (which include a
Boltzmann factor) are not corrected.

The J=7-6 line of CO shows emission over a broad velocity range (see Figure
3). The shape of this broad emission changes smoothly with position. This
gas is certainly warm and emits in high-J lines of CO as shown in the LWS
grating data of Sgr A$^*$ (see inset panel in Figure 3b). The J=3-2 line
shows several narrow features in absorption against the broad component
arising in the vicinity of Sgr A$^*$. These narrow absorptions (see Figure
3), at velocities of 0, -30 and -50 km$^{-1}$ have been also seen in many
other molecular and atomic species, and do correspond to cold gas in the
line of sight which produces the series of R and P lines up to J=7 in the
ro-vibrational band of CO (see Figure 1). Mapping with the CSO of both CO
lines indicates that the absorption features are present in all positions
where the broad emission is seen. These absorptions appear in all observed
positions (10x6 arcminutes). Hence, both broad emission and narrow lines
fill the ISO LWS and SWS beams.  Consequently, prior to the analysis of the
ro-vibrational absorption, the broad components due to the ices and, in the
case of \sgra\ to the warm CO gas in the Sgr A West region (Sutton et
al.~1990), had to be analyzed and removed.  In the following discussion we
will assume $x({\rm CO})=10^{-4}$ and $^{12}$CO$/^{13}{\rm CO} = 60$
(Langer and Wilson 1990).

\subsection{Modeling the Warm CO Gas in Sgr A West}

This warm component is clearly visible in Figure 1 and 3 (high-J CO lines;
inset panel in Figure 3b).  The CO ro-vibrational data (SWS) have been
modeled by a single component with $T_{\rm rot}=150\,$K, $\Delta v=
200\kms$, and $N({\rm CO}) = 1.0\per 10^{17}\cmi$.  The assumed linewidth
is suggested by our J=7-6 line data and by the LWS Fabry Perot (LWS-FP)
observations of some CO lines (not shown here).  The LWS-grating data
toward Sgr A$^*$ show CO pure rotational line emission from the $J=14$--13
up to the $J=21$--20 transitions (some of them are shown in the inset panel
of Figure 3b). Assuming that the emission fills the beam, and using a large
velocity gradient code and the collisional rates of Schinke et al.~(1985),
the observed intensities are reproduced, within 20--30\%, for $n({\rm
H_2})= 3\per 10^5\cc$, T$_{\rm K} = 250\,$K, $N({\rm CO})= 1.2\per
10^{17}\cmi$ with the assumed linewidth.  The corresponding excitation
temperatures vary between 160 to 180 K from the $J=14$--13 to the
$J=21$--20 rotational lines of CO, i.e., similar to the rotation
temperature derived from the SWS ro-vibrational absorption of CO. Note that
the region is rather complex and that the column density we derive
represents a beam averaged value. Column densities for individual clouds
could be larger and have to be derived using the corresponding filling
factor and the correct linewidths.  The values for n(H$_2$) and T$_K$ are
in excellent agreement with those derived by Sutton et al.~(1990) for the
individual warm clouds.  The region of GCS-3 is free of molecular emission
(Serabyn and G\"usten 1991), and thus no such warm gas was used in the
GCS-3 model.

\subsection{The CO and H$_2$O ices}

The CO, XCN, and \h2o ices shown in our spectra are among those that are
commonly found in the direction of young sources deeply embedded in
molecular clouds.  The CO and XCN ice absorption bands were modeled as
simple Gaussians; two Gaussians of different widths and central wavelengths
were used for the polar and non-polar CO ices (Chiar et al.~1998).  The
main goal of this fitting was to remove the CO-ice contribution from the
continuum, nevertheless we could derive $N($CO-ice$)\simeq 1.4\per 10^{17}$
and $\simeq 0.7\per 10^{17} \cmi$ for \sgra\ and GCS-3, respectively, with
$\sim 70\%$ apolar in \sgra, and $\sim 45\%$ apolar in GCS-3.  Our value of
$N$(CO-ice) in \sgra\ is twice that derived by McFadzean et al.~(1989) from
lower resolution spectroscopy.  As for the \h2o\ ice, Chiar et al.~(2000)
used the same SWS data discussed here, but considered both the $6\mic$ and
the $3\mic$ water ice features to determine $N$(H$_2$O-ice$) \simeq 1.2\per
10^{18}\cmi$ and $5\per 10^{17}\cmi$ for \sgra\ and GCS-3, respectively.

The high fraction of apolar CO-ice and the high $N$(CO)/$N$(\h2o) ratios,
together with the deep absorption of CO$_2$ ice (Whittet et al.~1997), are
typical of cold, quiescent environments (Chiar et al.~1998), indicating
that these ices cannot be located in the warm clouds in Sgr A West.

\subsection{The cold CO gas}

Due to the heavy saturation of the \aco\ ro-vibrational lines (opacities of
about 20--30), the physical parameters of the cold CO gas were determined
from the \bco\ and the adopted isotopic abundance ratio.  Three components
at the velocities indicated by the J=3-2 absorption features in Figure 3
were used in the model, with linewidths $\Delta v=12 \kms$ as measured in
our data. These linewidths agree well with the results shown in Figure 1 of
Sutton et al.~(1990).  The derived model parameters (column density and
rotation temperature) are the same for the three components. Model results
can be found in Table 1.  A single \c18o\ velocity component was introduced
in order to fit the weak $R(0)$ line of this isotope.  We also find that in
both lines of sight the cold gas alone is not sufficient to reproduce
completely the \aco\ absorption, and a component of higher kinetic
temperature, $T \simeq 40\,$K, and $N \simeq 1\%$ of the total, is
necessary to account for the highest observed $J$ lines.  This component
cannot arise from an incomplete removal of the warm gas in Sgr A West,
since it is present in both spectra, and should correspond to dense cloud
material ($n({\rm H_2}) > 5\per 10^4\cc$) of low CO column density.  We
estimate uncertainties of about a factor of 2 in the column densities.  The
volume density needed to pump the observed rotational levels of the CO
ground state is around $3\per 10^4\cc$ in the case of $T_{\rm rot} \simeq
10\,$K (Sgr A$^*$), decreasing to $\simeq 1000\cc$ at $\simeq 5\,$K
(GSC-3).  Adopting $A_V/N({\rm H_2}) \simeq 2\per 10^{-21}\,{\rm
mag}/\cmi$, the cold CO column densities imply visual extinction of 33 and
21 mag towards \sgra\ and GCS-3, respectively, generally consistent with
values in the literature.

\begin{table*}[htb] \begin{center}
\caption{~~~~~~~~~~~~~~~~~~~~~~~~~~~~~~~~~~~~~~~~~~~~~~~~~~~~~~~~
         Model results (T in K, N in $\cmi$) }
\begin{tabular}{lcc}
\tableline 
Quantity~~~~~~~~~~~~~~~~~~	 & \sgra~ & GCS-3~ \\
\tableline
\multicolumn{2}{c}{\bf warm gas} &\\
\tableline
T$_K$(CO \& \h2o)    &  250 &     \\
T$_{rot}$(CO)    &  $\simeq$150 &     \\
N($^{12}$CO) & $1.0\per 10^{17}$ &    \\
N(H$_2$O) &   $1.0\per 10^{15}$  &    \\
\tableline
\multicolumn{2}{c}{\bf cold gas} &   \\
\tableline
T$_{rot}$(CO)    &  6--12 &   5--10  \\
T$_{rot}$(\h2o)  &      8 &          \\
N($^{12}$CO) & $6.6\per 10^{18}$ &  $4.2\per 10^{18}$\\
N($^{13}$CO) & $1.1\per 10^{17}$ &  $0.7\per 10^{17}$\\
N(C$^{18}$O) & $1.3\per 10^{16}$ &  $1.7\per 10^{16}$\\ 
N(H$_2$O)    & $1.9\per 10^{16}$ &  $1.2\per 10^{16}$\\
\tableline 
\end{tabular} \end{center} \end{table*} \noindent

\subsection{The cold/warm H$_2$O gas}

In contrast to CO, the \h2o\ absorption at 6 $\mu$m is weak: 3\%\
($8\sigma$) in \sgra, and 2\%\ ($3\sigma$) in GCS-3 (not shown).  The three
features detected are the 0$_{00}$---1$_{11}$ (para), and the
1$_{01}$---1$_{10}$ and 1$_{01}$---2$_{12}$ (ortho) lines of the
$\nu_2=0\rightarrow1$ vibrational transition of water vapor (see Figure
1f).  In the \sgra\ data, the para transition is blended with the H$_2$
(0,0) S(6) line, which was removed by assuming an intensity of \onethird\
the average of the S(5) and S(7) lines.  The para line of H$_2$O is present
before the subtraction of the S(6) line of H$_2$, however, its intensity
is poorly determined.  Features B \& C in Figure 1f are ortho-H$_2$O while
A corresponds to para-H$_2$O.  Figure 1g shows the H$_2$O levels involved
in these transitions.  All three absorption features arise from the ground
para and ortho states to the first excited rotational levels of the bending
mode of water vapor.  The lack of other H$_2$O absorption features
indicates that all the water molecules are in the ground state (ortho and
para).  An ortho/para abundance ratio of 3 has been assumed.

The best fit to the H$_2$O absorption, assuming also the three velocity
components of CO, is obtained for $T_{\rm rot} \simeq 8\,$K, similar to the
temperatures deduced from CO.  While these \h2o\ lines are not very
sensitive to temperature in the 3--12 K range, an upper limit $T_{\rm rot}
\lesssim 12\,$K is obtained from the lack of absorption from the 1$_{10}$
ortho level which is 26.7 K above the ortho ground state (absorption from
other levels of para water is more difficult as the first excited level,
the 1$_{11}$, is 53.4 K above the para ground state).  The ro-vibrational
line opacities are $\simeq$ 1 and $\simeq 0.3$ for the ortho and para
species, respectively.  The column density of H$_2$O is $2\per 10^{16}\cmi$
with little dependence on the assumed $\Delta v$. This is due to the fact
that line opacity and spectral dilution effects have inverse dependency on
$\Delta$v for optically thin ($\tau$ below 1) lines.

The fact that the rotation temperature is low does not mean that the
kinetic temperature is also low.  While in the case of CO the inferred
densities are high enough to make $T_{\rm rot} \simeq T_{\rm kin}$, the
same is not true for \h2o. In order to determine what fraction of the
absorbing H$_2$O belongs to which temperature component, we used the LWS-FP
spectrum of the 1$_{01}$---2$_{12}$ pure rotational line of o-H$_2$O (in
the $\nu_2=0$ level) towards Sgr A$^*$ (see Figure 3b). It indicates that
the absorption arises from three different unresolved velocity components
at --20, 5 and $55\kms$.  The 1$_{10}$---2$_{21}$ line, on the other hand,
only shows significant absorption at $55 \kms$.  This cloud is probably the
one associated to Sgr A$\*$ itself and as for Sgr B2 (Cernicharo et al.,
1997) the analysis of the water vapor data is much more difficult due to
the role of dust emission/absorption in the excitation of this molecule.
Taking into account the absolute velocity accuracy of the LWS-FP, the --20
and $5\kms$ features could correspond to the cold NH$_3$ (Serabyn and
G\"usten 1986) and CO J=3-2 (see Figure 3a). The CO absorption at -50
kms$^{-1}$ is marginal in the LWS-FP data of H$_2$O.  For the physical
conditions of the cold gas, a minimum opacity of $\simeq 2$ with $\Delta
v=12\kms$ is needed to produce the observed absorption.  It corresponds to
$N({\rm H_2O}) > 2\per 10^{14}\cmi$.  However, with this column density the
contribution of the cold H$_2$O to the ro-vibrational absorption (SWS) will
be negligible.  Can we derive more precisely this column density?  An
answer can come from the different opacities of the pure rotational and
ro-vibrational lines of H$_2$O.

The line profiles of the water transitions differ remarkably from those of
the CO lines observed with the CSO telescope (see Figure 3a) and also with
the LWS-FP instrument (data not presented here).  The warm gas in Sgr A
West could also produce a broad H$_2$O absorption, and such absorption
would appear as a baseline effect in the reduced velocity coverage of the
LWS-FP spectra. Such a baseline has been removed from our data but it is
not possible to infer the possible contribution of the warm gas from these
data.  The grating data (see inset in Figure 3b), however, show an
absorption feature 13\%\ deep at the wavelength of the $1_{01}$---$2_{12}$
line (and very weak absorption is also found for the $1_{10}$---$2_{21}$
line), which would require $4\times$ the absorption produced by the narrow
lines of Figure 3b. Therefore, most of the absorption found in the grating
spectrum must correspond to a broad velocity component badly detected with
the LWS/FP. This broad absorption is also found in the 20 positions
observed with the LWS grating spectrometer around Sgr A$^*$.

The grating data suggest a line opacity $\simeq 1$ for a linewidth of $200
\kms$.  From the physical conditions we have derived for the warm gas, the
$1_{01}$---$2_{12}$ line opacity will be $\sim 1$ if $N({\rm H_2O}) =
10^{15}\cmi$. Hence, the H$_2$O abundance in the warm gas is $10^{-6}$,
similar to that found in the warm foreground gas of Sgr B2 
(Cernicharo et al.~1997a, Neufeld et al.~2000). However, this 
column density that explains well the absorption excess in the LWS grating 
data is insufficient to explain the SWS absorption. Hence, we have
to assume that the cold gas must have a larger column density to account
for the SWS data.

Removing the contribution from the warm gas, the total column density of
cold H$_2$O must be $1.9\per 10^{16}\cmi$ (three velocity components).  The
opacity and excitation temperature of the 1$_{01}$---2$_{12}$ rotational
line will be $\simeq 90$ and 7 K, respectively, for each one of the three
velocity components assumed in our models.  For GSC-3 we obtain a cold
H$_2$O column density of $\simeq 1.2\per 10^{16}\cmi$.  These results
suggest that the lines of sight intersect dense dark clouds where most of
the absorption is produced.  The \bco/H$_2$O abundance ratio would be
$\simeq 5.8$ in both cases, i.e, $^{12}{\rm CO/H_2O} \simeq 350$.  Hence,
$x({\rm H_2O})\simeq 3\per 10^{-7}$ and the water vapor gas/solid ratio is
$\sim 0.02$ in the quiescent regions of dark clouds along the line of sight
to the Galactic Center.

The water vapor abundance in the dark clouds intersecting the line of sight
of Sgr A$^*$ is much lower than that in the core of Sgr B2 and the shocked
regions of Orion (see introduction). However, the total gas+ice CO/H$_2$O
abundance ratio is $\simeq 5$. This result means that a large fraction of
the \h2o\ molecules is condensed on grains in these cold, quiescent regions
but also that the chemical mechanisms leading to its formation (either in
gas phase or in the grain surface) are very efficient and make H$_2$O the
most abundant molecular species after CO.

The ISO data corresponding to the pure rotational lines of H$_2$O towards
Sgr A$^*$ alone provide a lower limit for the water abundance of $\sim
10^{-8}$. However, when combining these data with the ro-vibrational H$_2$O
lines also observed by ISO a larger abundance, $\simeq$10$^{-7}$, is
found. With this abundance, the pure 2$_{12}$---1$_{01}$ and
1$_{10}$---1$_{01}$ rotational lines of H$_2$O in the ground vibrational
state will be extremely optically thick (see above), but the expected
emission from these lines will remain very low. The upper limit to the
water abundance derived through emission measurements by SWAS in the
direction of TMC1 ($7\per 10^{-8}$, Snell et al.~2000) is only of factor of
4 lower than the value we derive in the dark clouds towards Sgr A$^*$. If
we consider that in dark clouds important scattering effects have to be
taken into account in the radiative transfer of optically thick lines (see
Cernicharo and Gu\'elin, 1987), the upper limit derived by SWAS could be
slightly underestimated. Therefore, our results for Galactic dark clouds
towards Sgr A$^*$ and SWAS results for TMC1 would be in a relatively good
agreement. Finally, absorption measurements towards continuum background
sources at 556.9 GHz (1$_{01}$---1$_{10}$), and at 179.526 and
$269.273\mic$ ($1_{01}$---2$_{12}$ and $0_{00}$---1$_{11}$) with future
satellite missions such as FIRST, will provide a very efficient way to
derive the water vapor abundance in dark clouds and in the diffuse
interstellar medium.

\acknowledgements We thank Spanish DGES for this research under grants
PB96-0883 and ESP98-1351E. We thank E. Gonz\'alez-Alfonso for useful
comments. The CSO is supported by American NSF grant number AST-9980846.


\vskip 2cm

\begin{figure}[h]
\includegraphics{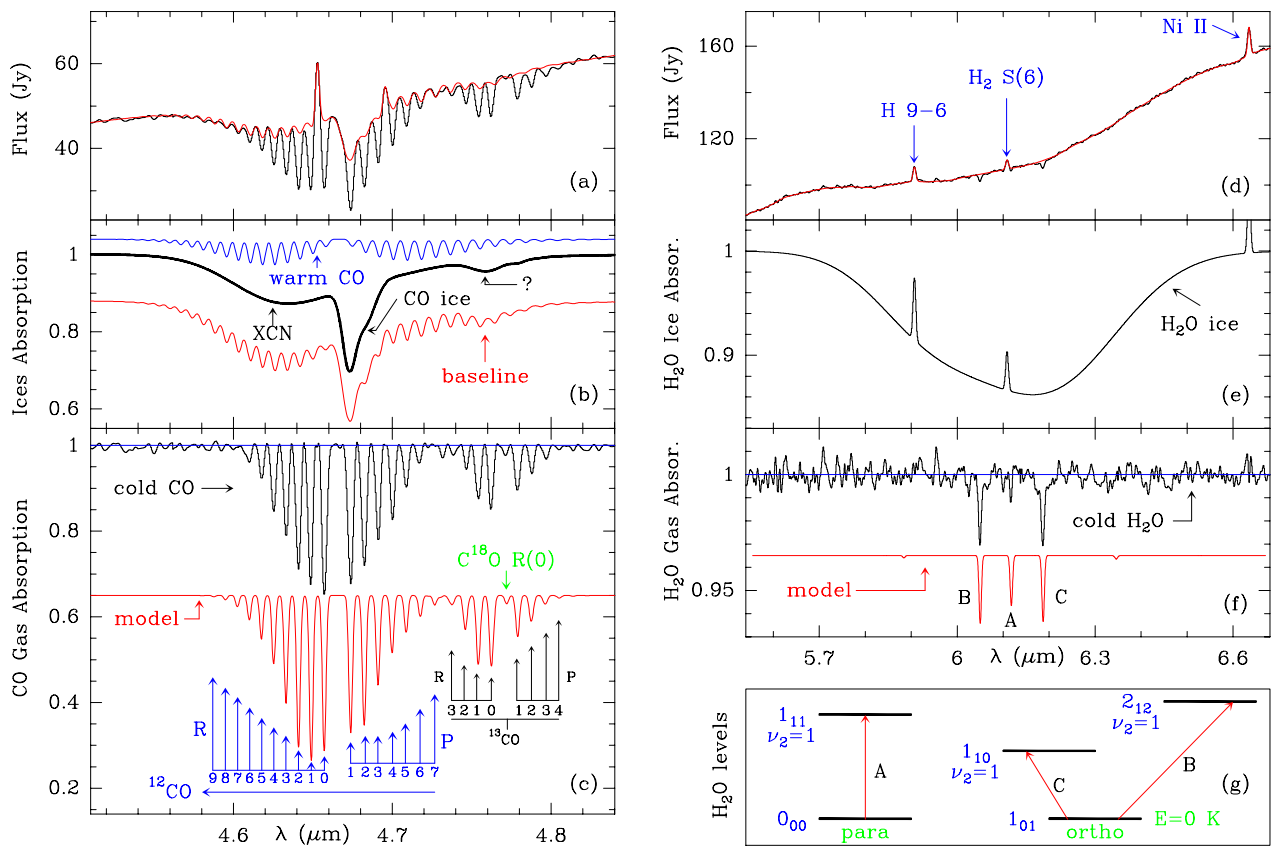}
\caption{ISO/SWS spectra of the stretching mode of CO {\bf (a,b,c)} and
the bending mode of water vapor {\bf (d,e,f)} in the direction of Sgr
A$^*$ (ISO observation number, ION, 46301102).  Panels {\bf (a)} and {\bf
(d)} show the observations and the adopted continuum (including ices);
panels {\bf (b)} and {\rm e} show the contribution of the ices and of the
warm CO; panels {\bf (c)} and {\bf (f)} show the normalized spectrum and
model spectrum (shifted for clarity), and panel {\bf (g)} shows the details
of the ortho- and para-\h2o transitions.  }
\end{figure}

\clearpage                                        

\begin{figure}[h]
\includegraphics{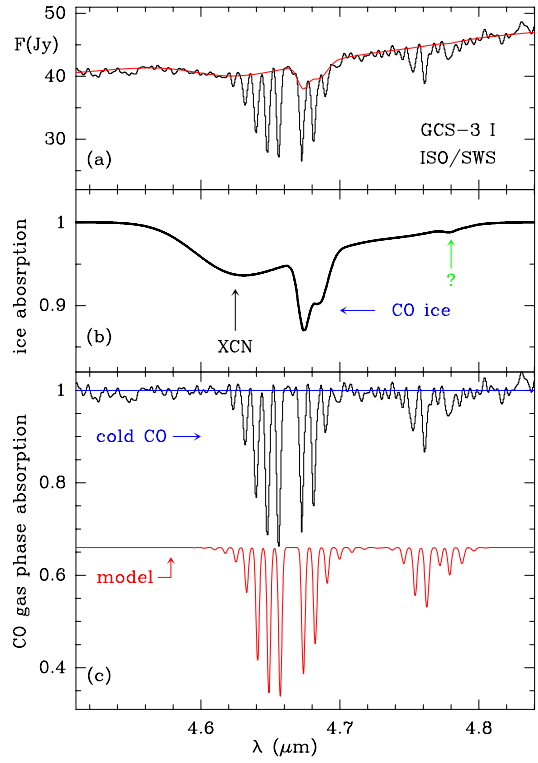}
\caption{CO spectrum and model for GCS-3 (ION 32701543).  Panel {\bf (a)}
shows the observations and the adopted continuum (including ices); panel
{\bf (b)} shows the contribution of the ices and of the warm CO; panel {\bf
(c)} shows the normalized spectrum and model spectrum (shifted for
clarity)}
\end{figure}

\clearpage

\begin{figure}[h]
\includegraphics{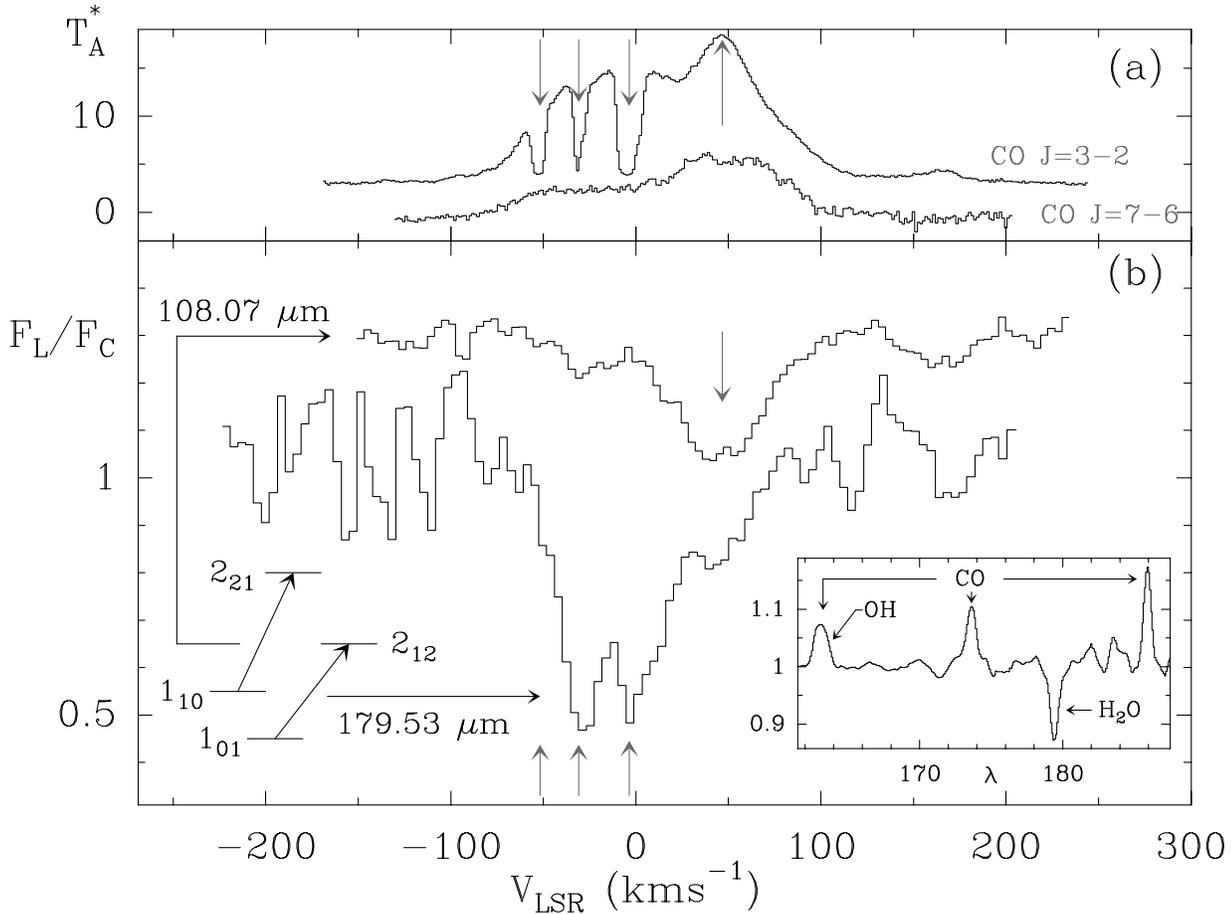}
\caption{ {\bf (a)} J=3-2 and J=7-6 lines of CO observed at
the CSO towards Sgr A$^*$. The downwards arrows indicate the central
position of the cold gas absorption features. The upwards arrows indicates
the velocity of the Sgr A West molecular cloud. {\bf (b)} LWS-FP spectra of
the 2$_{12}$-1${_01}$ and 2$_{21}$-1$_{10}$ lines of o-H$_2$O toward Sgr
A$^*$ (IONs 46300809 \& 46900837).  The drawing at bottom left indicates
the transitions levels involved.  The 1$_{01}$---2$_{12}$ line shows three
unresolved components at $v = 20$, 5 and $55\kms$ while in the other
transition only the $55\kms$ feature is present.  The uncertainty in the
absolute velocity $\simeq 10$--$15\kms$.  The inset panel shows the
LWS-grating spectrum (IONs 28701825, 32600211, 32702049, 49801004,
50501346, 67701010, 85302208 \& 82800298); the CO and H$_2$O lines are
indicated. The LWS-FP spectrum of o-H$_2$O has been shifted in velocity by
1/3 of the spectral resolution in order to match the velocity of the
absorption features of other molecular species including CO. The upwards
arrows indicate the velocities of the cold absorbing gas and the downwards
arrow that of the Sgr A$^*$ molecular cloud.}
\end{figure}

\enddocument